\documentclass[prd,superscriptaddress,amsfonts,amssymb,onecolumn,amsmath,showpacs]{revtex4-2}
\usepackage{bm}
\usepackage{amsfonts}
\usepackage{latexsym}
\usepackage[utf8]{inputenc}
\usepackage{graphicx}
\usepackage{amsmath}
\usepackage{palatino}
\usepackage{mathpazo}
\usepackage[british]{babel}
\usepackage{hhline}
\usepackage{multirow}
\usepackage{textcomp}
\usepackage{float}
\usepackage{booktabs}
\usepackage{dcolumn}
\usepackage{lipsum} 
\usepackage{mdframed}
\usepackage[]{mdframed}
\usepackage{hhline}
\usepackage{multirow}
\usepackage{ragged2e}
\usepackage{hyperref}
\hypersetup{colorlinks,citecolor=blue}
\hypersetup{colorlinks=true,linkcolor=red,filecolor=magenta,    urlcolor=cyan}
\usepackage{amsmath}
\usepackage{xcolor}
\usepackage{orcidlink}
\usepackage{epsfig}
\usepackage{caption}
\usepackage{subcaption}
\usepackage{commath}
\def\jnl@style{\it}
\def\aaref@jnl#1{{\jnl@style#1}}

\def\aaref@jnl#1{{\jnl@style#1}}

\def\aj{\aaref@jnl{AJ}}                   % Astronomical Journal
\def\apj{\aaref@jnl{ApJ}}                 % Astrophysical Journal
\def\apjl{\aaref@jnl{ApJ}}                % Astrophysical Journal, Letters
\def\apjs{\aaref@jnl{ApJS}}               % Astrophysical Journal, Supplement
\def\apss{\aaref@jnl{Ap\&SS}}             % Astrophysics and Space Science
\def\aap{\aaref@jnl{A\&A}}                % Astronomy and Astrophysics
\def\aapr{\aaref@jnl{A\&A~Rev.}}          % Astronomy and Astrophysics Reviews
\def\aaps{\aaref@jnl{A\&AS}}              % Astronomy and Astrophysics, Supplement
\def\mnras{\aaref@jnl{Mon.~Not.~Roy.~Astron.~Soc.}}             % Monthly Notices of the RAS
\def\prd{\aaref@jnl{Phys.~Rev.~D}}        % Physical Review D
\def\prc{\aaref@jnl{Phys.~Rev.~C}}  % Physical Review C
\def\prl{\aaref@jnl{Phys.~Rev.~Lett.}}    % Physical Review Letters
\def\qjras{\aaref@jnl{QJRAS}}             % Quarterly Journal of the RAS
\def\skytel{\aaref@jnl{S\&T}}             % Sky and Telescope
\def\ssr{\aaref@jnl{Space~Sci.~Rev.}}     % Space Science Reviews
\def\zap{\aaref@jnl{ZAp}}                 % Zeitschrift fuer Astrophysik
\def\nat{\aaref@jnl{Nature}}              % Nature
\def\aplett{\aaref@jnl{Astrophys.~Lett.}} % Astrophysics Letters
\def\apspr{\aaref@jnl{Astrophys.~Space~Phys.~Res.}} % Astrophysics Space Physics Research
\def\physrep{\aaref@jnl{Phys.~Rep.}}      % Physics Reports
\def\physscr{\aaref@jnl{Phys.~Scr}}       % Physica Scripta
\def\commat{\aaref@jnl{Comm.~Math.~Phys.}}              % Communications in Mathematical Physics
\def\science{\aaref@jnl{Science}}               % Science
\def\cqg{\aaref@jnl{Classical Quant.~Grav.}}            % Classical and Quantum Gravity
\def\jpcs{\aaref@jnl{JPCS}}                                     % Journal of Physics Conference Series
\def\ijmpd{\aaref@jnl{Int.~J.~Mod.~Phys.~D}}                    % International Journal of Modern Physics D
\def\grg{\aaref@jnl{Gen.~Relat.~Gravit.}}               % General Relativity and Gravitation
\def\rpp{\aaref@jnl{Rep.~Prog.~Phys.}}          % Reports on Progress in Physics
\def\npa{\aaref@jnl{Nucl.~Phys.~A}}        % Nuclear Physics A
\def\lrr{\aaref@jnl{Living Rev.~Rel.}}                   % Living reviews in relativity
\def\jcap{\aaref@jnl{J.~Cosmology Astropart.~Phys.}}    % Journal of cosmology and astroparticle physics
\def\rmp{\aaref@jnl{Rev.~Mod.~Phys.}}   %Reviews of modern physics
\def\epjc{\aaref@jnl{Eur.~Phys.~J.~C}} 
\def\plb{\aaref@jnl{~Phy.~Lett.~B}} 
\def\mpla{\aaref@jnl{Mod.~Phy.~Lett.~A}} 
\def\arxiv{\aaref@jnl{arxiv.org}}

\allowdisplaybreaks[1]
\begin{document}

\title{Dynamical System Analysis for Scalar Field Potential in Teleparallel Gravity}

\author{S. A. Kadam\orcidlink{0000-0002-2799-7870}}
\email{k.siddheshwar47@gmail.com}
\affiliation{Department of Mathematics, Birla Institute of Technology and Science-Pilani, Hyderabad Campus, Hyderabad-500078, India}

\author{Ananya Sahu\orcidlink{0009-0006-2284-9082}}
\email{ananyasahu0110@gmail.com}
\affiliation{Department of Physics, Indira Gandhi Institute of Technology, Sarang, Dhenkanal, Odisha-759146, India.}

\author{S. K. Tripathy \orcidlink{0000-0001-5154-2297}}
\email{tripathy\_sunil@rediffmail.com}
\affiliation{Department of Physics, Indira Gandhi Institute of Technology, Sarang, Dhenkanal, Odisha-759146, India.}

\author{B. Mishra\orcidlink{0000-0001-5527-3565}}
\email{bivu@hyderabad.bits-pilani.ac.in}
\affiliation{Department of Mathematics, Birla Institute of Technology and Science-Pilani, Hyderabad Campus, Hyderabad-500078, India}

\begin{abstract}
\textbf{Abstract:} In this paper, we have presented a power law cosmological model and its dynamical system analysis in $f(T,\phi)$ gravity, where $T$ is the torsion scalar and $\phi$ is the canonical scalar field. The two well-motivated forms of the non-minimal coupling function $F(\phi)$, the exponential form and the power law form, with exponential potential function, are investigated. The dynamical system analysis is performed by establishing the dimensionless dynamical variables, and the critical points were obtained. The evolution of standard density parameters is analysed for each case. The behaviour of the equation of state (EoS) and deceleration parameter show agreement with the result of cosmological observations. The model parameters are constrained using the existence and the stability conditions of the critical points describing different epochs of the evolution of the Universe. \\
{\bf Keywords:} Teleparallel Gravity, Scalar field, Potential Functions, Phase Space, Deceleration Parameter.
\end{abstract}

\maketitle

\section{Introduction}\label{Introduction}

Post supernovae era\cite{Riess:1998cb,Perlmutter:1998np}, there is a lot of debate on the late-time cosmic expansion phenomena of the Universe. Still, investigations are being pursued to obtain a conclusive explanation for this phenomenon. The exotic form of energy termed Dark energy (DE) believed to be responsible for this accelerating behaviour \cite{Hinshaw:2013, DIVALENTINO:2016}  accounted for around $68.3\%-70 \%$ of the mass-energy budget of the Universe \cite{Planck:2018vyg}. Based on Einstein's General Relativity (GR), the standard cosmology proposes that a cosmological constant $\Lambda$ is responsible for accelerated expansion. However, the $\Lambda$CDM (cold dark matter) model has severe fine-tuning problems with its energy scale \cite{Copeland:2006wr,Martin2012}. Furthermore, some tensions exist between cosmic observations and real measurements. The Planck data \cite{Planck:2018vyg,Planck:2015bue} shows a conflict with the weak lensing measurements and redshift surveys related to matter-energy density $(\Omega_{m})$ and structure growth rate ($f_{\sigma_8}$). Also, the $H_{0}$ tension \cite{DiValentino2021}, which describes the differences in the CMB and the direct local distance ladder measurements \cite{Riess:2011,Riess:2016}. Hence exploring new physics beyond the standard cosmological model may play a crucial role in studying the cause of cosmic acceleration. \\

Two main approaches are discussed in the literature to explain the late-time accelerated expansion of the universe. The first method involves adding an exotic fluid with negative pressure to the matter part of the Einstein field equations. This has led to the development of various models that use different scalar fields, such as quintessence \cite{Tsujikawa:2013fta,Chiba:1999ka}, k-essence \cite{Armendariz-Picon:2000ulo}, phantom \cite{roy2018dynamical}, tachyon fields \cite{Otalora:2013dsa,Sen_2002}, quintom \cite{Cai:2009zp}. The DE can also be explained by the second way of obtaining modification into the geometric part of the Einstein field equations. One of the possible ways in this direction is to replace the Levi-Civita connection \cite{misner1973gravitation} with the Weizenb$\ddot{o}$ck connection \cite{Weitzenbock1923,Bahamonde:2021gfp}. This new formalism is dynamically equivalent to GR, called the teleparallel equivalent of GR (TEGR) \cite{Hayashi:1979qx}. This was introduced by Einstein as an alternative to GR \cite{Linder:2010py}. Although the GR and TEGR produce the same field equations, but were constructed differently. GR uses the symmetric Levi--Civita \cite{misner1973gravitation} connection to define the curvature and produces zero torsion. In contrast, teleparallel gravity (TG) uses a nonsymmetric Weitzenböck connection \cite{Weitzenbock1923,Bahamonde:2021gfp} with no curvature but only torsion. The TEGR lagrangian consists of the torsion scalar $T$ term, obtained through the contraction of the torsion tensor. In other words, GR uses the Ricci scalar $R$ to describe the gravitational effects, while the TEGR uses torsion instead. One of the interesting extensions to TEGR is the $f(T)$ gravity \cite{Ferraro:2006jd, Cai:2015emx,Bengochea:2008gz,Duchaniya:2022rqu}, which involves an arbitrary function of the torsion scalar $T$. This model has been extensively studied in the literature, particularly to explain the accelerated expansion of the universe through the torsion approach \cite{Basilakos:2018arq,Ferraro:2006jd, Cai:2015emx,Bengochea:2008gz,Duchaniya:2022rqu,DUCHANIYA2024f(T)}. \\

Similar to the modifications made in GR with the scalar field \cite{Capozziello:2011et}, the same has been introduced in the TEGR \cite{Cai:2015emx,Bahamonde:2017ize}. It is important to note that although TEGR and GR are dynamically equivalent, a scalar-torsion theory with non-minimal coupling is not equivalent to its curvature-based counterpart. A scalar-torsion theory with a non-minimal coupling term, $\xi \phi^2 T$, where $\phi$ is the dynamical scalar field, was initially applied to DE in Ref. \cite{Geng:2011aj,Geng_2012}. These modifications can be more generalised by introducing the general form of the non-minimally coupled scalar field form $F(\phi)G(T)$. These generalisations are assumed to be an extension of $f(T)$ gravity in terms of the gravitational sector \cite{Ferraro:2008ey,Linder:2010py}. Motivated by the study of scaling solutions presented in Ref, \cite{Uzan1999,Amendola1999}, in TEGR, the $f(T,\phi)$ gravity \cite{Gonzalez-Espinoza:2021mwr,Gonzalezreconstruction2021,Gonzalez-Espinoza:2020jss} can be considered. The $f(T,\phi)$ gravity describes the modification of the non-minimal coupling function $F(\phi)$ to the function $G(T)$ of torsion scalar $T$ in a more general way. The covariant formulation of scalar torsion theory has been explored in Ref. \cite{Hohmann:2018rwf}. In this theory, the Lagrangian is dependent on both the canonical scalar field and the general function of torsion scalar $T$. The scalar torsion $f(T, \phi)$ gravity model has a crucial role in the comprehensive study of the cosmological scenario. For example, it is involved in the stability of scalar perturbations in the presence of matter fluid as in Ref. \cite{Gonzalez-Espinoza:2021mwr}, as well as the study of cosmological singularities as in Ref. \cite{trivedi2023cosmological}. Additionally, Ref. \cite{Gonzalezreconstruction2021} analyses the reconstruction of $f(T, \phi)$ gravity to determine the forms of non-minimally coupling function of the scalar field $F(\phi)$ and the scalar field potential $V(\phi)$ by using the power law form of the torsion scalar $T$. \\

The dynamical system techniques are valuable tools for investigating the complete asymptotic behavior of a given cosmological model \cite{Kadam:2022lgq,Kadamdynamicalftb,duchaniya2023dynamical,DUCHANIYA2024101402,KADAM2024Aop} and also it helps to bypass the difficulty of solving non-linear cosmological equations. These tools describe the global dynamics of the Universe by analyzing the local asymptotic behavior of critical points of the system and relating them to the main cosmological epochs of the universe\cite{Paliathanasis:2017flf}. This approach also plays a vital role in constraining the cosmological viable range of the model parameters, which helps in the study of the theory to describe important cosmological epochs \cite{Santos2018}.  The study of analyzing dynamical systems for both power law and logarithmic forms of the torsion scalar is made in detail in Refs.\cite{samaddar2023qualitative,Gonzalez-Espinoza:2020jss,duchaniya2023dynamical} for the exponential form of the coupling function. In the reconstruction analysis stated in Ref. \cite{Gonzalezreconstruction2021}, the power law form of the torsion scalar $T$ produces both the exponential and power law forms for the non-minimally coupled scalar field function $F(\phi)$ and the scalar field potential $V(\phi)$ in $f (T,\phi)$ gravity. The motivation behind the present study is to examine the dynamics of the Universe for both the exponential and power law forms of the non-minimally scalar field coupling function $F(\phi)$, coupled with the exponential scalar field potential $V(\phi)$. The late-time cosmic acceleration behaviour can be studied through the value of the equation of state (EoS) parameter, which we intend to find with the dimensionless variables described in the dynamical system. The corresponding model parameters would be constrained for both scalar field coupling functions.\\

The paper is organised in several sections: In Section \ref{Backgroundformalism}, discusses the teleparallel gravity formalism for $f(T,\phi)$ gravity. In Section \ref{dynamicalsystemanalysis}, the autonomous dynamical system analysis has been formulated. The dynamical system analysis has been presented in two sub-sections: Subsection \ref{Exponentialcouplingfun} covers the exponential coupling function, and Subsection \ref{Powerlawcouplingfun} covers the power law potential coupling function. Finally, in Section \ref{conclusion}, we summarize the results.

\section{\texorpdfstring{$f(T,\phi)$}{} gravity field equations}\label{Backgroundformalism}

The action for $f(T,\phi)$ gravity is \cite{Gonzalez-Espinoza:2021mwr},
\begin{equation}\label{ActionEq}
S =\int d^{4}xe[f(T,\phi)+P(\phi)X]+S_{r}+S_{m}
\end{equation}
where the determinant of the tetrad field, $e = \det[e^A_{\mu}] = \sqrt{-g}$; and the action for matter and radiation are respectively denoted as $S_{m}$ and $S_{r}$. We know that GR can also be represented in the framework of teleparallel gravity, known as TEGR. This is because, in place of the metric tensor of GR, the tetrad and spin connection pair can be treated as the dynamical variables in TEGR. One can have the local relationship between metric tensor $(g_{\mu \nu})$, tetrad field $(e^{A}_{\mu}, A,\mu = 0,1,2,3)$ and the Minkowski tangent space metric $(\eta_{AB})$ as
\begin{equation}\label{TetradEq1}
g_{\mu \nu}=\eta_{AB} e_{\mu}^{A} e_{\nu}^{B}, \\
\end{equation}
where $\eta_{AB}=(-1,1,1,1)$. Also, $e^{\mu}_Ae^B_{\mu}=\delta_A^B$, $i.e.$ tetrad satisfies the orthogonality condition. We consider non-minimal coupled scalar-torsion gravity models with the coupling function $f(T, \phi)$, where $\phi$ and $T$ respectively denotes the scalar field and torsion scalar. In the second term of the action, the kinetic term of the field, $X= -\partial_{\mu} \phi \partial^{\mu} \phi/2$. The torsion scalar, $T = S_{\theta}^{\mu \nu} T_{\mu \nu}^{\theta}$, where $S_{\theta}^{\mu \nu}$ represents the superpotential and $T_{\mu \nu}^{\theta}$ be the torsion tensor and can be respectively expressed as,
\begin{eqnarray}
S_{\theta}^{~~\mu \nu}&\equiv&\frac{1}{2}(K^{\mu \nu}_{~~~\theta}+\delta^{\mu}_{\theta}T^{\alpha \nu}_{~~~\alpha}-\delta^{\nu}_{\theta}T^{\alpha \mu}_{~~~\alpha})\label{Superpotential} \\
T_{\mu \nu}^{\theta} &=& e^{\theta}_{A}\partial_{\mu} e^{A}_{\nu}-e^{\theta}_{A}\partial_{\nu}e^{A}_{\mu}+e^{\theta}_{A} \omega^{A}_{B\mu}e^{B}_{\nu}-e^{\theta}_{A} \omega^{A}_{B\nu}e^{B}_{\mu}\label{Torsiontenosr} \,. 
\end{eqnarray} 

In Eq. \eqref{Superpotential}, the contortion tensor takes the form, $K^{\mu \nu}_{~~~\theta}\equiv \frac{1}{2}(T^{\nu \mu}_{~~~\theta}+T_{\theta}^{~~\mu \nu}-T^{\mu \nu}_{~~~\theta})$. To note, there exists some special frame in which the spin connection vanishes, known as the Weitzenb$\ddot{o}$ck gauge. Varying the action Eq.\eqref{ActionEq} with respect to the tetrad $(e^{A}_{\mu})$, the gravitational field equation can be obtained  \cite{Hohmann:2018rwf}.  The relation between curvature and torsion scalar can be established using $(T = -R + 2e^{-1}\partial_{\mu}(eT^{\alpha\mu}_{~~\alpha}))$. For the  homogeneous and isotropic flat Friedmann-Lema\^{i}tre-Robertson-Walker (FLRW) space-time,
\begin{equation}\label{FLRWSpacetime}
ds^{2}=-dt^{2}+a^{2}(t)[dx^2+dy^2+dz^2]\,,
\end{equation}
the tertad becomes $e^{A}_{\mu} = diag(1,a,a,a)$, where $a(t)$ represents the expansion rate. Varying the action in Eq.~\eqref{ActionEq} with respect to the tetrad field and the scalar field $\phi$, we can obtain the field equations of $f(T,\phi)$ gravity as,
\begin{eqnarray}
f(T,\phi)-P(\phi)X-2Tf,_{T}&=&\rho_{m}+\rho_{r}\,, \label{FE1}\\
f(T,\phi)+P(\phi)X-2Tf,_{T}-4\dot{H}f,_{T}-4H\dot{f},_{T} &=& -p_{r}\,,\label{FE2}\\
-P,_{\phi}X-3P(\phi)H\dot{\phi}-P(\phi)\ddot{\phi}+f,_{\phi}&=&0\label{KleinGordon1}\,. 
\end{eqnarray}

For brevity, we represent $f\equiv f(T,\phi)$ and $f_{, T}=\frac{\partial f}{\partial T}$, $H\equiv\frac{\dot{a}}{a}$ be the Hubble parameter with an over dot represents the derivative with respect to the cosmic time $t$.  $\rho_{m}$, $\rho_{r}$, and $p_{r}$ respectively represent the matter-energy density, radiation energy density, and radiation pressure. The torsion scalar for the flat FLRW space-time becomes $T=6H^{2}$. We refer the non-minimal coupling function $f(T,\phi)$ in the form \cite{Hohmann:2018rwf}

\begin{equation}\label{Generalmodel}
f(T,\phi)=-\frac{T}{2\kappa^{2}}-F(\phi)G(T)-V(\phi)\,,
\end{equation}
where $V(\phi)$ is the scalar potential, $F(\phi)$ is an non-minimally coupling function of the scalar field and $G(T)$ is the arbitrary function of torsion scalar $T$. The condition for matter dominated era $\omega_{m}=\frac{p_{\rm m}}{\rho_{\rm m}} = 0$, and radiation dominated era $\omega_{r}=\frac{p_{r}}{\rho_{\rm r}} = \frac{1}{3}$ have been imposed. Now, Eqs.~\eqref{FE1}--\eqref{KleinGordon1} reduce to
\begin{align}
\frac{3}{\kappa^2}H^2& =P(\phi)X+V(\phi)-2TG_{,T} F(\phi)+G(T) F(\phi)+\rho_m+\rho_r\label{FEmr1}\\
-\frac{2}{\kappa^2}\dot{H}& =2P(\phi)X+4\dot{H} G_{,T}F(\phi) +4HG_{,TT} \dot{T} F(\phi) +4HG_{,T} \dot{F}+\rho_m+\frac{4}{3}\rho_r\label{FEmr2}\\
&P(\phi)\ddot{\phi}+3P(\phi)H\dot{\phi} +P_{,\phi} X +G(T) F_{,\phi} +V_{,\phi}=0\label{KleinGordon2}
\end{align}

The Friedmann Eqs.~\eqref{FEmr1}--\eqref{FEmr2} can also be written as
\begin{eqnarray}
\frac{3}{\kappa^{2}}H^{2}=\rho_{m}+\rho_{r}+\rho_{DE}\label{FriedmannEq1}\,, \\
-\frac{2}{\kappa^{2}}\dot{H}=\rho_{m}+\frac{4}{3}\rho_{r}+\rho_{DE}+p_{DE}\label{FriedmannEq2}\,, 
\end{eqnarray}
so that one can retrieve the energy density [Using Eq.~\eqref{FEmr1} and Eq.~\eqref{FriedmannEq1}] and pressure [Using Eq.~\eqref{FEmr2} and Eq.~\eqref{FriedmannEq2}] for the DE sector respectively as,
\begin{align}
\rho_{DE} &= P(\phi)X+V(\phi)+(G(T)-2TG_{,T}) F(\phi),\label{FEDE1}\\
p_{DE} &= P(\phi) X-V(\phi)-G(T) F(\phi)+2TG_{,T} F(\phi) +8TG_{,TT} F(\phi)\dot{H}+4 G_{,T} F(\phi)\dot{H} +4HG_{,T} F_{,\phi} \dot{\phi}.\label{FEDE2}
\end{align}
We consider $P(\phi)$ = 1 and the potential energy, $V(\phi)=V_{0}e^{-\kappa\lambda\phi}$, \cite{samaddar2023qualitative,Gonzalez-Espinoza:2020jss,duchaniya2023dynamical} where $\lambda$ is a constant. The fluid equation can be written as
 \begin{align}\label{ConservationEq}
 \dot{\rho_r}+4H\rho_r=0\,,\nonumber\\
 \dot{\rho_m}+3H \rho_{m}=0\,,\nonumber\\
\dot{\rho_{DE}}+3H(\rho_{DE}+p_{DE})=0.
 \end{align}
The standard density parameters for the matter, radiation and the DE sector are respectively  $\Omega_m=\frac{\kappa^2\rho_m}{3H^2}$, $\Omega_r=\frac{\kappa^2\rho_r}{3H^2}$ and $\Omega_{DE}=\frac{\kappa^2\rho_{DE}}{3H^2}$ such that
 \begin{equation}\label{ConstrainEq}
 \Omega_m+\Omega_r+\Omega_{DE}=1.
 \end{equation}

The late-time cosmic acceleration issue is a recent phenomenon in modern cosmology. Several cosmological models are available in the literature to find reasons for this strange behavior of the Universe. However, to study the stability of these models, dynamical system analysis has been an effective tool \cite{Bahamonde:2017ize,Wu:2010xk}. Hence, in this study, we aim to study the dynamical system analysis with the torsion-based gravitational theory coupled with the scalar field. To do so, we need a certain form of $G(T)$ to be incorporated in Eqs. \eqref{FEDE1}--\eqref{FEDE2} and hence we consider the well known forms of $G(T)$ in the following section.

\section{The Dynamical system analysis With Power Law Model }\label{dynamicalsystemanalysis}
%********************************
%********************************
We consider $G(T)= \alpha \left(-T\right)^{\beta}$, where $\alpha$ and $\beta$ are model parameters \cite{Mirza_2017,briffa2023,Bengochea:2008gz}. Subsequently,  Eqs. \eqref{FEDE1}--\eqref{FEDE2} reduce to
 
 \begin{align}
 \rho_{DE} &= X+V(\phi)+ F(\phi)\left[\alpha\left(1-2\beta\right)\right]\left(-T\right)^{\beta},\label{FEDErho}\\ 
 p_{DE} &= X-V(\phi)- F(\phi) \left[\alpha\left(1-2\beta\right)\right](-T)^{\beta}+8\dot{H} F(\phi) T \alpha \beta (\beta-1)(-T)^{\beta-2}\nonumber\\ 
  & \,\,\, +4 (-\alpha \beta) (-T)^{\beta-1} F(\phi) \dot{H}+4H(-\alpha \beta (-T)^{\beta-1}) F_{,\phi} \dot{\phi}\,\label{FEDEp}.
 \end{align}
 
 The scalar field Klein-Gordon equation presented in Eq. \eqref{KleinGordon2} can take the following form, 
 \begin{equation}
 \ddot{\phi}+3H\dot{\phi}+\alpha(-T)^{\beta} F_{,\phi}+V_{,\phi}(\phi)=0,\label{KleinGordonM1}
 \end{equation}

The following dimensionless phase space variables are given to obtain the autonomous dynamical system, 
\begin{align} 
x &= \frac{\kappa\dot{\phi}}{\sqrt{6}H}, \quad
y = \frac{\kappa \sqrt{V}}{\sqrt{3}H}, \quad
u = \frac{\kappa^2 F(\phi)\left[\alpha (1-2\beta)(-T)^{\beta}\right]}{3H^2},\ \quad \rho = \frac{\kappa\sqrt{\rho_r}}{\sqrt{3}H},\nonumber\\
\lambda &= -\frac{V_{,\phi}(\phi)}{\kappa V(\phi)} \quad,
\Gamma = \frac{V(\phi)V_{,\phi\phi}}{V_{,\phi}^2(\phi)},\quad
\sigma = -\frac{F_{,\phi}(\phi)}{\kappa F(\phi)}  \quad,  \Theta = \frac{F(\phi)F_{,\phi\phi}}{F_{,\phi}^2(\phi)}.\label{dynamicalvariables}
\end{align}
The density parameters [Eq. \eqref{ConstrainEq}] in terms of dimensionless variables are,
\begin{eqnarray} \label{densityeqs}
\Omega_r & = & \rho^2, \nonumber\\
\Omega_m & = & 1-x^2-y^2-u-\rho^2,\nonumber\\
\Omega_{DE} & = & x^2+y^2+u.
\end{eqnarray}
The deceleration parameter $(q)$, which shows the accelerating or decelerating behaviour of the Universe can be obtained as 
\begin{equation}\label{decelerationEq}
q=-1-\frac{\dot{H}}{H^2} =-1-\frac{u \left(2 \beta  \left(\sqrt{6} \lambda  x-3\right)+3\right)+(2 \beta -1) \left(\rho ^2+3 x^2-3 y^2+3\right)}{2 (2 \beta -1) (\beta  u-1)}, 
\end{equation}

where the following can be obtained in terms of dimensionless variables using Eqs. \eqref{FEDE1}--\eqref{FEDE2},

\begin{equation} \label{eq25}
\frac{\dot{H}}{H^2} = \frac{u \left(2 \beta  \left(\sqrt{6} \lambda  x-3\right)+3\right)+(2 \beta -1) \left(\rho ^2+3 x^2-3 y^2+3\right)}{2 (2 \beta -1) (\beta  u-1)}.
\end{equation}

The EoS parameters for the total and the DE sector, respectively, can be obtained as,

\begin{eqnarray}
\omega_{tot} &=&-1-\frac{u \left(2 \beta  \left(\sqrt{6} \lambda  x-3\right)+3\right)+(2 \beta -1) \left(\rho ^2+3 x^2-3 y^2+3\right)}{3 (2 \beta -1) (\beta  u-1)}, \label{eq26}\\
\omega_{DE} &=& \frac{u \left(\beta  \left(-2 \beta  \left(\rho ^2+3\right)+\rho ^2-2 \sqrt{6} \lambda  x+9\right)-3\right)-3 (2 \beta -1) (x-y) (x+y)}{3 (2 \beta -1) (\beta  u-1) \left(u+x^2+y^2\right)}.\label{eq27}
\end{eqnarray}

For $N=ln a$, $a$ is the scale factor; consequently, we have the autonomous dynamical system for the power law model as, 
\begin{eqnarray}
\frac{dx}{dN} &=& -3 x+\frac{\sqrt{6} \lambda  y^2}{2}+\frac{\sqrt{6} \sigma  u}{2-4 \beta}+ \left(-\frac{x \left(u \left(2 \beta  \left(\sqrt{6} \lambda  x-3\right)+3\right)+(2 \beta -1) \left(\rho ^2+3 x^2-3 y^2+3\right)\right)}{2 (2 \beta -1) (\beta  u-1)}\right),\label{dynamicaleq1}\\
\frac{dy}{dN} &=& \frac{1}{2} y \left(\left(-\sqrt{6}\right) \lambda  x-\frac{u \left(2 \beta  \left(\sqrt{6} \lambda  x-3\right)+3\right)+(2 \beta -1) \left(\rho ^2+3 x^2-3 y^2+3\right)}{(2 \beta -1) (\beta  u-1)}\right), \label{dynamicaleq2}\\
\frac{du}{dN} &=&-\sqrt{6} \sigma  u x+\frac{(\beta -1) u \left(u \left(2 \beta  \left(\sqrt{6} \lambda  x-3\right)+3\right)+(2 \beta -1) \left(\rho ^2+3 x^2-3 y^2+3\right)\right)}{(2 \beta -1) (\beta  u-1)}, \label{dynamicaleq3}\\
\frac{d\rho}{dN} &=& -\rho  \left(\frac{u \left(2 \beta  \left(\sqrt{6} \lambda  x-3\right)+3\right)+(2 \beta -1) \left(\rho ^2+3 x^2-3 y^2+3\right)}{2 (2 \beta -1) (\beta  u-1)}+2\right), \label{dynamicaleq4}\\
\frac{d\lambda}{dN} &=& -\sqrt{6}\left(\Gamma-1\right)\lambda^2 x, \label{dynamicaleq5}\\
\frac{d\sigma}{dN} &=& -\sqrt{6}\left(\Theta-1\right)\lambda^2 x.\label{dynamicaleq6}
\end{eqnarray}

Now, we will solve the equations $\frac{dx}{dN}=0$,  $\frac{dy}{dN}=0$, $\frac{du}{dN}=0$, $\frac{d\rho}{dN}=0$, $\frac{d\lambda}{dN}=0$ and $\frac{d\sigma}{dN}=0$ to derive the critical points of the autonomous dynamical system to study the dynamical features of the evolution of Universe. The stability characteristics are classified using the conditions as (a) The stable node: The sign of all the eigenvalues are negative; (b) The unstable node: all eigenvalues are positive; (c) Saddle point: one, two, or three of the four eigenvalues are positive, and the others are negative; (d) Stable spiral: the determinant of the Jacobian matrix is negative, and the the real part of all eigenvalues is negative.  The attractor points are the critical points that fall in the stability condition for stable nodes or stable spirals, which can be reached through cosmological evolution. Another type of critical point with zero eigenvalues is the non-hyperbolic critical point \cite{coley2003dynamical}. In this case, the linear stability theory fails to determine the stability of the critical point. Hence, we applied the specific condition in which if the number of vanishing eigenvalues is equal to the dimension of the set of critical points, it is normally hyperbolic. The stability can be analyzed by deriving the conditions for which the non-vanishing eigenvalues are negative.  While studying the cosmic scenarios, it is imperative to find solutions where the energy density of the scalar field matches with the background energy density $(\frac{\rho_{DE}}{\rho_{m}}=\tau)$, where $\tau$ is a non--zero constant. The cosmological solutions satisfying this property are known as the scaling solutions\cite{Otalora:2013dsa,Otalora:2013tba}. In this study, we also analyse the scaling solutions with different non-minimally coupled scalar field functions. We have analyse two cases with two forms of $F(\phi)$.

\subsection{\texorpdfstring{$V(\phi)=V_0e^{-\lambda \kappa \phi}, \,  F(\phi)=F_0e^{-\eta  \kappa \phi} $}{}}\label{Exponentialcouplingfun}

The exponential form of the coupling function is widely studied in the literature to study the dynamical system analysis in scalar field models\cite{Gonzalez-Espinoza:2020jss,roy2018dynamical}. In this case, using Eq. \eqref{dynamicalvariables}, we have $\Gamma=1$ and $\Theta=1$. Hence, the autonomous dynamical system presented in  Eq. \eqref{dynamicaleq1}--Eq.\eqref{dynamicaleq6} reduce to four equations with four independent variables $x,\, y,\, u,\, \rho$. The critical points for this system are presented in Table-\ref{modelIcriticalpoints} along with the values of $\omega_{tot}$ and the standard density parameters $\Omega_{m}$, $\Omega_{r}$, $\Omega_{DE}$. The stability of the critical point is analyzed using the eigenvalues of the Jacobian matrix at each critical point. For this case, the eigenvalues, along with the stability conditions, are presented in Table-\ref{modelIeigenvalues}.

\begin{figure}[H]
    \centering
\includegraphics[width=58mm]{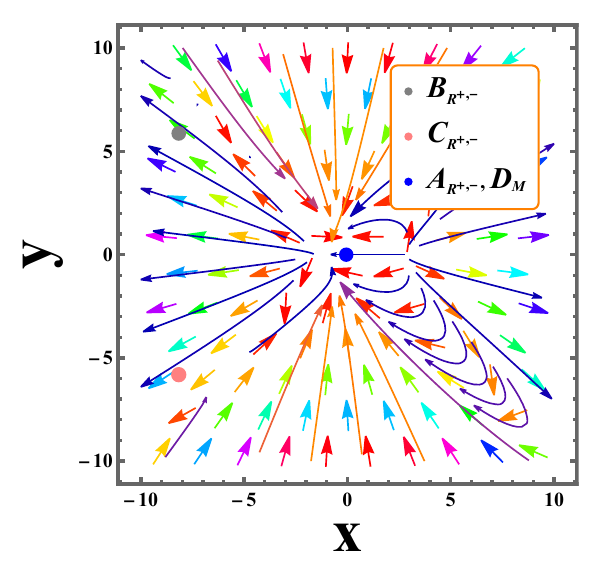}
\includegraphics[width=58mm]{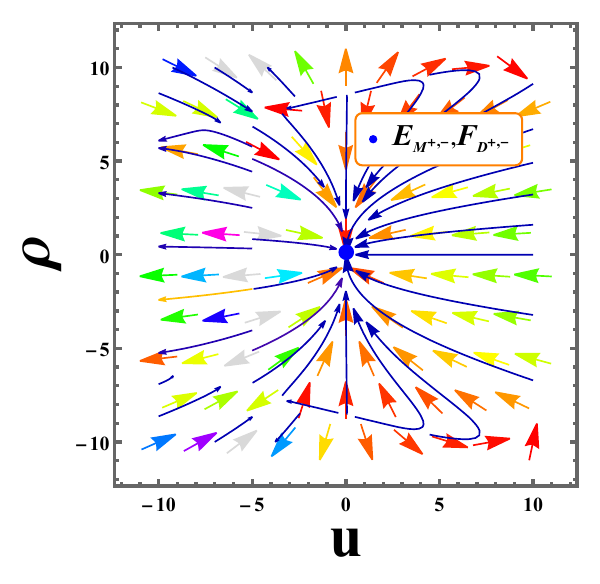}
\includegraphics[width=58mm]{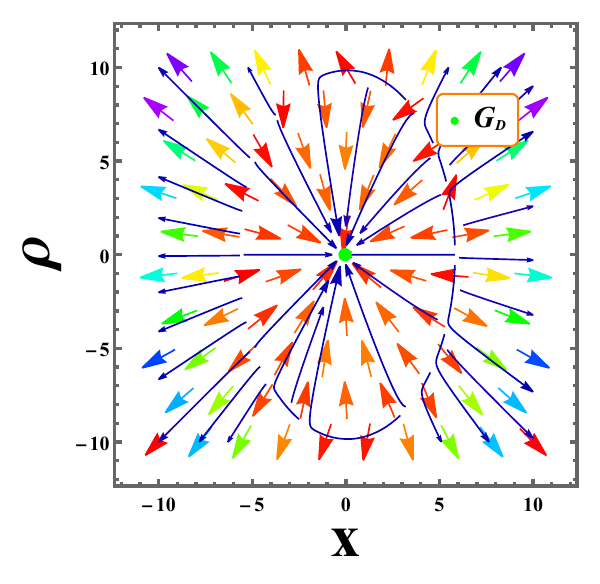}
    \caption{ 2-D Phase Portraits at $x=0,\, y=-4.9,\, u=8,\, \rho=4.5,\, \beta=-0.2, \beta=0$ for ($G_{D}$),  \, $ \sigma=-0.30, \, \lambda=-0.2$ for [Model-\ref{Exponentialcouplingfun}].}
    \label{phasespace2dm1}
\end{figure}

\begin{figure}[H]
\centering
\includegraphics[width=70mm]{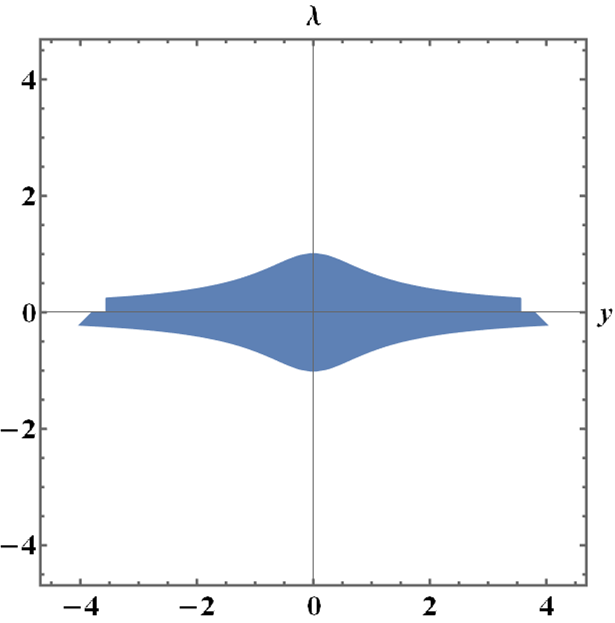}
\includegraphics[width=70mm]{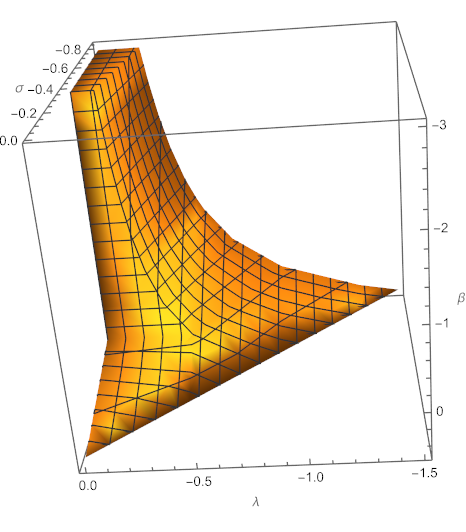}
\caption{ Region plot for stability condition of critical point $G_{D}$ and 3-D Region plot for common range of stability conditions for critical points for model parameters for$-\frac{\sqrt{3}}{2}<\sigma<0\&\&\left(\beta \leq \frac{1}{2}\land -\frac{\sigma }{\beta _{10}-1}<\lambda<0\right)$ for [Model-\ref{Exponentialcouplingfun}]. .} \label{regionplotm1}
\end{figure}

\begin{table}[H]
     % title of Table
    \centering % used for centering table
    \begin{tabular}{|c |c |c |c| c| c| c|} % centered columns (5 columns)
    \hline\hline %inserts double horizontal lines
    \parbox[c][0.9cm]{1.3cm}{{Name}
    }& $ \{ x_{c}, \, y_{c}, \, u_{c}, \, \rho_{c} \} $ & {Existence Condition} &  {$\omega_{tot}$}& $\Omega_{r}$& $\Omega_{m}$& $\Omega_{DE}$\\ [0.5ex] % inserts table %headin$g$
    \hline\hline % inserts single horizontal line
    \parbox[c][1.3cm]{1.3cm}{$A_{R^{\pm}}$ } &$\{ 0, 0, 0, \pm 1 \}$ & $2 \beta -1\neq 0$ &  $\frac{1}{3}$& $1$& $0$ & $0$ \\
    \hline
    \parbox[c][1.3cm]{1.3cm}{$B_{R^{\pm}}$ } & $\{\frac{2 \sqrt{\frac{2}{3}}}{\lambda}, \,  \frac{2}{\sqrt{3}\lambda}, \, 0 ,  \, \pm \sqrt{1-\frac{4}{\lambda^2}}\}$ & $\lambda\ne 0\, ,2 \beta -1\neq 0 $&  $\frac{1}{3}$& $1-\frac{4}{\lambda^2}$ & $0$ & $\frac{4}{\lambda^2}$ \\
    \hline
    \parbox[c][1.3cm]{1.3cm}{$C_{R^{\pm}}$ } & $\{\frac{2 \sqrt{\frac{2}{3}}}{\lambda}, \,  -\frac{2}{\sqrt{3}\lambda}, \, 0 ,  \, \pm \sqrt{1-\frac{4}{\lambda^2}}\}$ & $\lambda\ne 0\, ,2 \beta -1\neq 0 $&  $\frac{1}{3}$& $1-\frac{4}{\lambda^2}$ & $0$ & $\frac{4}{\lambda^2}$\\
    \hline
   \parbox[c][1.3cm]{1.3cm}{$D_{M}$ } &  $\{0, \, 0, \, 0, \, 0 \}$ & $2 \beta -1\neq 0$ &  $0$& $0$ & $1$ & $0$\\
   \hline
   \parbox[c][1.3cm]{1.3cm}{$E_{M^{\pm}}$} &   $\{\frac{\sqrt{\frac{3}{2}}}{\lambda},\pm\frac{\sqrt{\frac{3}{2}}}{\lambda},0,0\}$ & $2 \beta -1\neq 0$ &  $0$& $0$ & $1-\frac{3}{\lambda^2}$ & $\frac{3}{\lambda^2}$\\
   \hline
   \parbox[c][1.3cm]{1.3cm}{$F_{D^{\pm}}$} & $\{\frac{\lambda}{\sqrt{6}},\pm\sqrt{1-\frac{\lambda^2}{6}},0,0 \}$ & $2 \beta -1\neq 0$ &  $-1+\frac{\lambda^2}{3} $& $0$ & $0$ & $1$\\
   \hline
 \parbox[c][1.3cm]{1.3cm}{$G_{D}$} &  $\{0, y,1-y^2,0 \}$ & $\begin{tabular}{@{}c@{}}$\beta=0\,, \sigma =\frac{\lambda  y^2}{y^2-1}\,,$\\  $y^2-1\neq 0$\end{tabular}$& $-1$ & $0$ & $0$ & $1$\\
 \hline
 \end{tabular}
\caption{Critical points with the existence condition for {\bf[Model-\ref{Exponentialcouplingfun}}].}
% is used to refer to this table in the text
\label{modelIcriticalpoints}
\end{table}
%%%%%%%%%%%%%%%%%%%%%%%%%%%%%%%%%%%%%%%%%%%
\begin{table}[H]
\small\addtolength{\tabcolsep}{-7pt}
     % title of Table
    \centering % used for centering table
    \begin{tabular}{|c |c |c |c| c|} % centered columns (5 columns)
    \hline\hline %inserts double horizontal lines
    \parbox[c][0.9cm]{1.3cm}{{C. P.}
    }& Eigenvalues & Stability Conditions \\ [0.5ex] % inserts table %headin$g$
    \hline\hline % inserts single horizontal line
    \parbox[c][1.3cm]{1.3cm}{$A_{R^{\pm}}$ } & $\Big[-1,1,2,-4 (\beta -1)\Big]$ & Saddle at $\beta >1$\\
    \hline
    \parbox[c][1.3cm]{1.3cm}{$B_{R^{\pm}}$ } & $\Big[1,-\chi_{1}-\frac{1}{2},\frac{1}{2} \left(\chi_{1}-1\right),-4 \beta-\frac{4 \sigma}{\lambda}+4\Big]$ & Saddle at $\left(\beta>\frac{1}{2} \left(2-\sigma\right)\land \left(-\frac{8}{\sqrt{15}}\leq \lambda <-2\lor 2<\lambda\leq \frac{8}{\sqrt{15}}\right)\right)$\\
    \hline
   \parbox[c][1.3cm]{1.3cm}{$C_{R^{\pm}}$ } & $\Big[1,-\chi_{1}-\frac{1}{2},\frac{1}{2} \left(\chi_{1}-1\right),-4 \beta-\frac{4 \sigma}{\lambda}+4\Big]$  & Saddle at $\left(\beta>\frac{1}{2} \left(2-\sigma\right)\land \left(-\frac{8}{\sqrt{15}}\leq \lambda <-2\lor 2<\lambda\leq \frac{8}{\sqrt{15}}\right)\right)$\\
   \hline
   \parbox[c][1.3cm]{1.3cm}{$D_{M}$} &  $\Big[\frac{1-2 \beta}{2 \left(2 \beta-1\right)},-\frac{3}{2},\frac{3}{2},-3 \left(\beta-1\right)\Big]$ &  Saddle at $\beta >1$  \\
   \hline
    \parbox[c][1.3cm]{1.3cm}{$E_{M^{\pm}}$} &  $\begin{tabular}{@{}c@{}} $\Big[-\frac{1}{2},-\frac{3 \sqrt{-\left(1-2 \beta\right){}^2 \lambda^2 \left(7 \lambda^2-24\right)}}{4 \left(2 \beta-1\right) \lambda^2}-\frac{3}{4},$\\$\frac{3}{4} \left(\frac{\sqrt{-\left(1-2 \beta\right){}^2 \lambda^2 \left(7 \lambda^2-24\right)}}{\left(2 \beta-1\right) \lambda^2}-1\right),-3 \beta-\frac{3 \sigma}{\lambda}+3\Big]$\end{tabular}$ &$\begin{tabular}{@{}c@{}} Stable at  \\ $2 \sigma<\lambda\leq 2 \sqrt{\frac{6}{7}}\land \beta >\frac{\lambda-\sigma}{\lambda}$ \end{tabular}$ \\
 \hline
 \parbox[c][1.3cm]{1.3cm}{$F_{D^{\pm}}$} &  $\Big[\frac{1}{2} \left(\lambda^2-6\right),\frac{1}{2} \left(\lambda^2-4\right),\lambda^2-3,-\beta \lambda^2-\lambda \sigma +\lambda^2\Big]$ & $\begin{tabular}{@{}c@{}}  \\Stable at \\$\beta\in \mathbb{R}\land \Big(\left(-\sqrt{3}<\lambda<0\land \sigma<\lambda-\beta \lambda \right)$\\$
 \lor \left(0<\lambda<\sqrt{3}\land \sigma>\lambda-\beta \lambda \right)\Big)$\end{tabular}$  \\
 \hline
 \parbox[c][1.3cm]{1.3cm}{$G_{D}$} & $\begin{tabular}{@{}c@{}} $\Big[-3,-2,-\frac{\sqrt{3} \sqrt{\left(y^2-1\right) \left(\left(4 \lambda^2+3\right) y^2-3\right)}}{2 \left(y^2-1\right)}-\frac{3}{2},$\\$\frac{1}{2} \left(\frac{\sqrt{3} \sqrt{\left(y^2-1\right) \left(\left(4 \lambda^2+3\right) y^2-3\right)}}{y^2-1}-3\right)\Big]$\end{tabular}$ & $\begin{tabular}{@{}c@{}} Stable at\\ $\lambda\in \mathbb{R}\land \lambda\neq 0 $\\$\land \left(-\sqrt{3} \sqrt{\frac{1}{4 \lambda^2+3}}\leq y<0\lor 0<y\leq \sqrt{3} \sqrt{\frac{1}{4 \lambda^2+3}}\right)$\end{tabular}$ \\
 \hline
\end{tabular}
    \caption{Eigenvalues and the stability conditions corresponding to each critical point for {\bf[Model-\ref{Exponentialcouplingfun}}], where $\chi_1=\frac{\sqrt{-\left(1-2 \beta \right){}^2 \lambda^2 \left(15 \lambda^2-64\right)}}{2 \left(2 \beta-1\right) \lambda^2}$.}
    % is used to refer to this table in the text
    \label{modelIeigenvalues}
\end{table}
%%%%%
Detailed analysis of each of the critical points are given below:
\begin{itemize}
\item \textbf{Radiation-Dominated Critical Points :} The critical points $A_{R_{\pm}},B_{R_{\pm}}$, and $C_{R_{\pm}}$ are the points in the radiation-dominated phase. The critical point $A_{R_{\pm}}$ with $\Omega_{r}=1$ represents the standard radiation-dominated era of the Universe evolution. The eigenvalues of the Jacobian matrix at this critical point contain both positive and negative signs hence it is an unstable saddle critical point. The value of $\omega_{tot}=\frac{1}{3}$ at this critical point. The other two critical points $B_{R_{\pm}}$, and $C_{R_{\pm}}$ are the scaling radiation-dominated solutions with $\Omega_{DE}=\frac{4}{\lambda^2}$. These critical points represent the non-standard radiation-dominated era with $\Omega_{r}=1-\frac{4}{\lambda^2}$. The eigenvalues at these critical points can be observed from Table-\ref{modelIeigenvalues}, which contain one positive. The other eigenvalues can take negative values in the range $\beta>\frac{1}{2} \left(2-\sigma\right)\land \left(-\frac{8}{\sqrt{15}}\leq \lambda<-2\lor 2<\lambda\leq \frac{8}{\sqrt{15}}\right)$ hence are saddle in this range and is unstable. The behaviour of phase space trajectories at these critical points can be analysed from Fig.--\ref{phasespace2dm1}. The plot is for  $u=8,\, \rho=4.5,\, \beta=-0.2, \,  \sigma=-0.30, \, \lambda=-0.2$ and it has been observed that the trajectories are moving away from these critical points hence the unstable behaviour can be confirmed.

\item{\textbf{Matter-Dominated Critical Points :}} The critical points $ D_{M}, E_{M^{\pm}}$ are the matter-dominated critical points. The critical point $D_{M}$ is the standard matter-dominated critical point with $\Omega_{m}=1$. This critical point is a saddle critical point with positive and negative eigenvalues hence it is unstable. The same can be observed from Table--\ref{modelIIeigenvalues}. In Fig.--\ref{phasespace2dm1}, the phase space trajectories are moving away from this critical point, confirming the unstable behaviour. The other critical point $E_{M^{\pm}}$ represents non-standard matter-dominated critical point with $\Omega_{M}=1-\frac{3}{\lambda^2}$. This critical point is a matter-dominated scaling solution, and the value of $\Omega_{DE}$ at this critical point is $\frac{3}{\lambda^2}$. From the eigenvalues presented in Table--\ref{modelIeigenvalues}, one can observe that this critical point is stable at $2 \sigma<\lambda\leq 2 \sqrt{\frac{6}{7}}\land \beta >\frac{\lambda-\sigma}{\lambda}$. At this critical point, the attracting phase space trajectories can be observed from Fig.--\ref{phasespace2dm1}. The value of $\omega_{tot}=0$, hence both the critical point can not describe the accelerating phase of the Universe.

\item{\textbf{DE-Dominated Critical Points :}} The critical points $F_{D^{\pm}}$ and $G_{D}$ are representing the standard DE-dominated era with $\Omega_{DE}=1$. The critical point $F_{D}$ is a stable DE-dominated solution with the stability range $\beta\in \mathbb{R}\land \Big(\left(-\sqrt{3}<\lambda<0\land \sigma<\lambda-\beta \lambda \right)
 \lor \left(0<\lambda<\sqrt{3}\land \sigma>\lambda-\beta \lambda \right)\Big)$. This critical point is a stable attractor, and the attracting behaviour of the phase space trajectories can be studied from Fig.--\ref{phasespace2dm1}. The value of $\omega_{tot}=-1+\frac{\lambda^2}{3}$ and can explain accelerated expansion of the Universe at $-\sqrt{2}<\lambda<\sqrt{2}$. This critical point will exist at the condition $2\beta-1\ne0$. The critical point $G_{D}$ is a de-Sitter solution with $\omega_{tot}=-1$. This solution exists at $\beta=0$. From Table--\ref{modelIeigenvalues} this critical point is stable and show stability at $\lambda\in \mathbb{R}\land \lambda\neq 0 \land \left(-\sqrt{3} \sqrt{\frac{1}{4 \lambda^2+3}}\leq y<0\lor 0<y\leq \sqrt{3} \sqrt{\frac{1}{4 \lambda^2+3}}\right)$. To get better clarity the 2-D plot for this range in terms of the dynamical variable $y$ and $\lambda$ is plotted in Fig--\ref{regionplotm1}.  The phase space trajectories at this critical point show the attracting behaviour, and hence, this critical point is a stable attractor. One important result is that, the power law form with exponential potentials is capable of describing different phases of the evolution of the Universe. The common parametric range at which the critical point $E_{M^{\pm}}, \, F_{D^{\pm}},\, G_{D}$ is stable at $-\frac{\sqrt{3}}{2}<\sigma<0\&\&\left(\beta \leq \frac{1}{2}\land -\frac{\sigma }{\beta _{10}-1}<\lambda<0\right)$ and for better visualisation the same has been shown graphically in Fig.--\ref{regionplotm1}.
\end{itemize} 

It is to note that the common stability range for critical points $E_{M^{\pm}}, F_{D^{\pm}}, G_{D}$ is $-\frac{\sqrt{3}}{2}<\sigma<0\&\&\left(\beta \leq \frac{1}{2}\land -\frac{\sigma }{\beta _{10}-1}<\lambda<0\right)$.

\begin{figure}[H]
    \centering
\includegraphics[width=70mm]{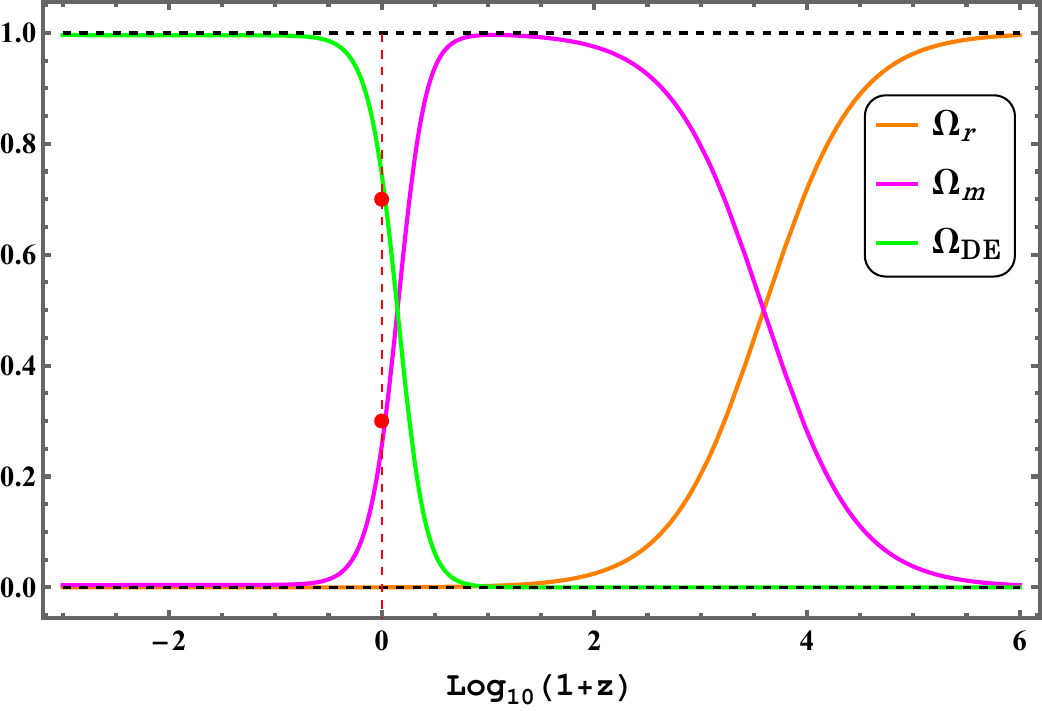}    \includegraphics[width=70mm]{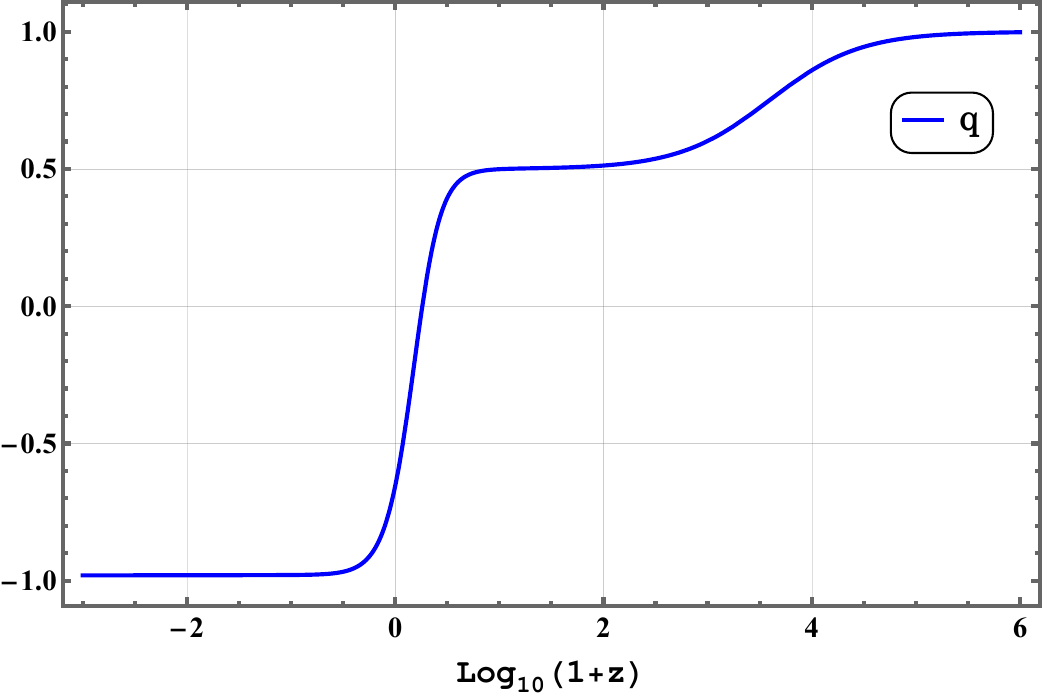}
    \caption{ {\bf Left Panel}--Density parameter for radiation $(\Omega_r)$, matter $(\Omega_m)$, DE $(\Omega_{DE})$ in redshift. {\bf Right Panel}--Deceleration parameter $(q)$ in redshift for $\beta=-0.2, \,  \sigma=-0.30, \, \lambda=-0.2$ for the initial conditions $x_0=10^{-8.89}, \, y_0=10^{-2.89}, \, u_0=10^{-5.96}, \, \rho_0= 10^{-0.9}$ for {\bf[Model-\ref{Exponentialcouplingfun}}]. } \label{evolutionm1}
\end{figure}

\begin{figure}[H]
    \centering
\includegraphics[width=70mm]{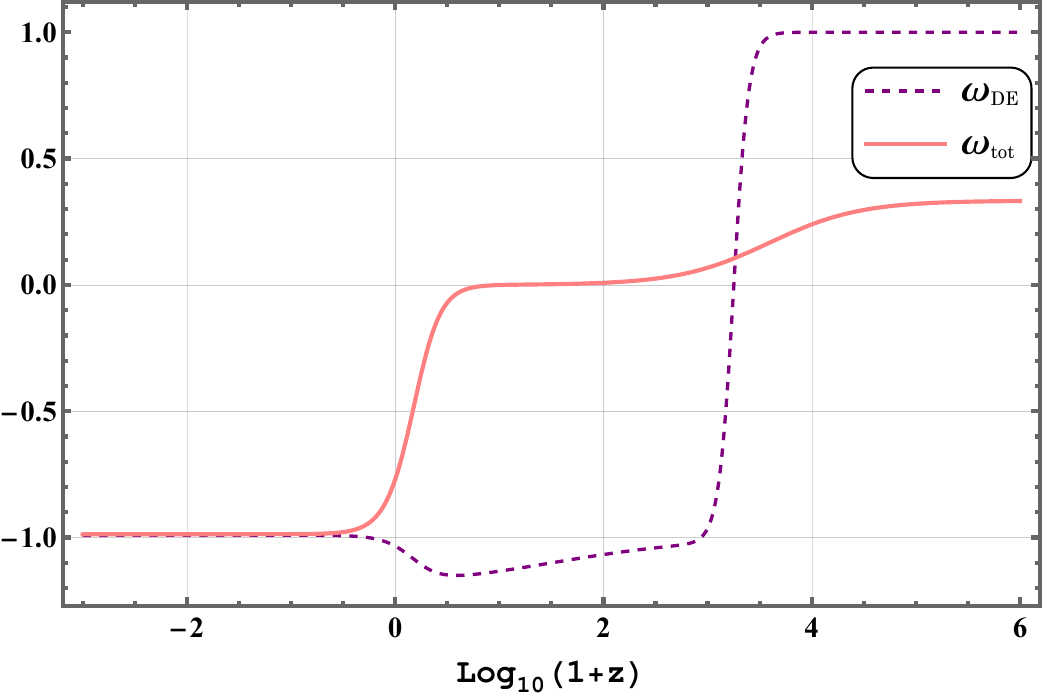}
    \caption{ The DE EoS $(\omega_{DE})$ parameter and total EoS $(\omega_{tot})$ parameter in redshift for $\beta=-0.2, \,  \sigma=-0.30, \, \lambda=-0.2$ for the initial conditions $x_0=10^{-8.89}, \, y_0=10^{-2.89}, \, u_0=10^{-5.96}, \, \rho_0= 10^{-0.9}$ for {\bf[Model-\ref{Exponentialcouplingfun}}].} \label{Eosplotm1}
\end{figure}

%%%%%%%
In Fig.--\ref{evolutionm1}, we have shown the evolutionary behaviour of the energy density pertaining to the matter, radiation, DE phase, and the deceleration parameter. It can be observed that the redshift of radiation
matter equality is around $z \approx 3387$, and the transition to the  accelerated phase at $z \approx 0.62$ (Ref. Fig.--\ref{evolutionm1}), which is compatible with the
 $\Lambda$CDM value. From the evolution plots of standard density parameters, we observe that at present, $\Omega_{m}\approx 0.3$, which agrees with the Planck observation results \cite{Planck:2018vyg}. The value of the evolution of DE at the present time is observed to be $\Omega_{DE} \approx 0.7$ \cite{Kowalski_2008}. The deceleration parameter shows decreasing behaviour from early to late. It lies in the negative region in the present and the late time, hence capable of describing the accelerating behaviour of the Universe. The present value of $q \approx -0.614^{+0.002}_{-0.002}$ and is compatible with the observational study made in \cite{Capozziellomnras}. In Fig.--\ref{Eosplotm1}, the total equation of state $\omega_{tot}$ and the equation of state of DE $\omega_{DE}$ is plotted. We observe that at late times, both the curves approaches to $\Lambda$CDM behaviour, though $\omega_{DE}$ shows phantom behaviour just before the present time.

%%%%%%%%%%%%%%%%%%%%%%%%%%%%%%%%%%%%%%%%%%%%%%%%%%%%%%%%%%%%%%%%%%%%%%%%%%%%%%%%%%%%%%%%%%%%%%%%%%%%%%%%%%%%%%%%%%%%%%%%%%%%%
\subsection{\texorpdfstring{$V(\phi)=V_{0}e^{-\lambda \kappa \phi}, F(\phi)=F_{0} \phi^n$}{}}\label{Powerlawcouplingfun}
Using Eq. \eqref{dynamicalvariables}, the exponential and the power law potential choice gives $\Gamma=1$ and $\Theta=\frac{n-1}{n}$ respectively and the dynamical system will have five independent variables $\left(x,\, y,\, u,\, \rho,\, \sigma\right)$ variables. Apart from the system of equations [Eq--(\ref{dynamicaleq1}--\ref{dynamicaleq4})], the following will be an additional equation in the system,
\begin{eqnarray}
\frac{d\sigma}{dN} &=& \frac{\sqrt{6} \sigma ^2 x}{n}.\label{DynamicalEq7}
\end{eqnarray}

 The critical points and the value of $\omega_{tot}, \Omega_{r}, \Omega_{m}$ and $\Omega_{DE}$ for this case are presented in Table-\ref{modelIIcriticalpoints}. The stability of these critical can be analysed using the eigenvalues. The eigenvalues with the stability conditions at these critical points are presented in Table-\ref{modelIIeigenvalues}. 
\begin{table}[H]
     % title of Table
    \centering % used for centering table
    \begin{tabular}{|c |c |c |c| c| c| c|} % centered columns (5 columns)
    \hline\hline %inserts double horizontal lines
    \parbox[c][0.9cm]{1.3cm}{{Name}
    }& $ \{ x_{c}, \, y_{c}, \, u_{c}, \, \rho_{c}, \, \sigma_{c} \} $ & {Existence Condition} &  \textbf{$\omega_{tot}$}& $\Omega_{r}$& $\Omega_{m}$& $\Omega_{DE}$\\ [0.5ex] % inserts table %headin$g$
    \hline\hline % inserts single horizontal line
    \parbox[c][1.3cm]{1.3cm}{$a_{R^{\pm}}$ } &$\{ 0, 0, 0, \pm 1, \, \sigma \}$ & $2 \beta  n-n\neq 0$ &  $\frac{1}{3}$& $1$& $0$ & $0$ \\
    \hline
    \parbox[c][1.3cm]{1.3cm}{$b_{R^{\pm}}$ } & $\{\frac{2 \sqrt{\frac{2}{3}}}{\lambda}, \,  \frac{2}{\sqrt{3}\lambda}, \, 0 ,  \, \pm \sqrt{1-\frac{4}{\lambda^2}}\, , 0\}$ & $ \lambda\ne 0\, ,2 \beta  n-n\neq 0 $&  $\frac{1}{3}$& $1-\frac{4}{\lambda^2}$ & $0$ & $\frac{4}{\lambda^2}$ \\
    \hline
    \parbox[c][1.3cm]{1.3cm}{$c_{R^{\pm}}$ } & $\{\frac{2 \sqrt{\frac{2}{3}}}{\lambda}, \,  -\frac{2}{\sqrt{3}\lambda}, \, 0 ,  \, \pm \sqrt{1-\frac{4}{\lambda^2}}\, , 0\}$ & $\lambda\ne 0\, , 2 \beta  n-n\neq 0$&  $\frac{1}{3}$& $1-\frac{4}{\lambda^2}$ & $0$ & $\frac{4}{\lambda^2}$ \\
    \hline
   \parbox[c][1.3cm]{1.3cm}{$d_{M}$ } &  $\{0, \, 0, \, 0, \, 0, \, \sigma \}$ & $2 \beta  n-n\neq 0$ &  $0$& $0$ & $1$ & $0$\\
   \hline
   \parbox[c][1.3cm]{1.3cm}{$e_{M}^{\pm}$} &   $\{\frac{\sqrt{\frac{3}{2}}}{\lambda }, \, {\pm} \frac{\sqrt{\frac{3}{2}}}{\lambda }, \, 0, \, 0, \, 0\}$ & $2 \beta  n-n\neq 0, \, \beta=2$ &  $0$& $0$ & $1-\frac{3}{\lambda^2}$ & $\frac{3}{\lambda^2}$\\
   \hline
   \parbox[c][1.3cm]{1.3cm}{$f_{D}{\pm}$} & $\{\frac{\lambda}{\sqrt{6}}, \,\pm \sqrt{1-\frac{\lambda^2}{6}} \, 0, \, 0,\, 0\}$ & $2 \beta  n-n\neq 0, \, \beta=2$ &  $-1+\frac{\lambda^2}{3}$& $0$ & $0$ & $1$\\
 \hline
 \parbox[c][1.3cm]{1.3cm}{$g_{D}$} &  $\{ 0, \, \frac{1}{2}, \, \frac{3}{4}, \, 0, \, \frac{-\lambda}{3} \}$ & $\beta=0, \, n=2$ &  $-1$ & $0$ & $0$ & $1$\\
 \hline
 \end{tabular}
\caption{Critical points with the existence condition for {\bf[Model-\ref{Powerlawcouplingfun}}].}
% is used to refer to this table in the text
\label{modelIIcriticalpoints}
\end{table}
%%%%%%

\begin{table}[H]
     % title of Table
    \centering % used for centering table
    \begin{tabular}{|c |c |c |c| c|} % centered columns (5 columns)
    \hline\hline %inserts double horizontal lines
    \parbox[c][0.9cm]{1.3cm}{\textbf{C. P.}
    }& Eigenvalues & Stability Conditions \\ [0.5ex] % inserts table %headin$g$
    \hline\hline % inserts single horizontal line
    \parbox[c][1.3cm]{1.3cm}{$a_{R^{\pm}}$ } & $\Big[0,-1,1,2,-4 (\beta -1)\Big]$ & Saddle at $\beta >1$ \\
    \hline
    \parbox[c][2.0cm]{1.3cm}{$b_{R^{\pm}}$ } & $\begin{tabular}{@{}c@{}}$\Big[0,1,-4 \left(\beta-1\right),-\frac{\sqrt{-\left(1-2 \beta\right){}^2 \lambda^2 \left(15 \lambda^2-64\right)}}{2 \left(2 \beta-1\right) \lambda^2}-\frac{1}{2},$\\$\frac{1}{2} \left(\frac{\sqrt{-\left(1-2 \beta\right){}^2 \lambda^2 \left(15 \lambda^2-64\right)}}{\left(2 \beta -1\right) \lambda^2}-1\right)\Big]$\end{tabular}$ & Saddle at $\beta>1\land \left(-\frac{8}{\sqrt{15}}\leq \lambda<-2\lor 2<\lambda\leq \frac{8}{\sqrt{15}}\right)$\\
    \hline
   \parbox[c][1.3cm]{1.3cm}{$c_{R^{\pm}}$ } & $\begin{tabular}{@{}c@{}}$\Big[0,1,-4 \left(\beta-1\right),-\frac{\sqrt{-\left(1-2 \beta\right){}^2 \lambda^2 \left(15 \lambda^2-64\right)}}{2 \left(2 \beta -1\right) \lambda^2}-\frac{1}{2},$\\$\frac{1}{2} \left(\frac{\sqrt{-\left(1-2 \beta\right){}^2 \lambda^2 \left(15 \lambda^2-64\right)}}{\left(2 \beta-1\right) \lambda ^2}-1\right)\Big]$\end{tabular}$ & Saddle at $\beta>1\land \left(-\frac{8}{\sqrt{15}}\leq \lambda<-2\lor 2<\lambda\leq \frac{8}{\sqrt{15}}\right)$\\
   \hline
   \parbox[c][1.3cm]{1.3cm}{$d_{M}$} &  $\Big[0,-\frac{3}{2},-\frac{1}{2},\frac{3}{2},-3 (\beta -1)\Big]$ &  Saddle at $\beta >1$  \\
   \hline
 \parbox[c][1.3cm]{1.3cm}{$e_{M}{\pm}$} &  $\Big[0,-3,-\frac{1}{2},\frac{3 \left(-\lambda^2-\sqrt{24 \lambda^2-7 \lambda ^4}\right)}{4 \lambda^2},\frac{3 \left(\sqrt{24 \lambda^2-7 \lambda^4}-\lambda^2\right)}{4 \lambda^2}\Big]$ &$\begin{tabular}{@{}c@{}} Stable at \\ $-2 \sqrt{\frac{6}{7}}\leq \lambda<-\sqrt{3}\lor \sqrt{3}<\lambda\leq 2 \sqrt{\frac{6}{7}}$ \end{tabular}$\\
 \hline
 \parbox[c][1.3cm]{1.3cm}{$f_{D}{\pm}$} &  $\Big[0,-\lambda^2,\frac{1}{2} \left(\lambda^2-6\right),\frac{1}{2} \left(\lambda^2-4\right),\lambda^2-3\Big]$ &  $\begin{tabular}{@{}c@{}} \\Stable at \\ $\sqrt{3}<\lambda <0\lor 0<\lambda <\sqrt{3}$\end{tabular}$  \\
 \hline
 \parbox[c][1.3cm]{1.3cm}{$g_{D}$} & $\begin{tabular}{@{}c@{}} $\Big[0,-3,-2,\frac{1}{4} \left(-\sqrt{2} \sqrt{18-7 \lambda^2}-6\right),\frac{1}{4} \left(\sqrt{2} \sqrt{18-7 \lambda^2}-6\right)\Big]$ \end{tabular}$& $\begin{tabular}{@{}c@{}} Stable at\\ $-3 \sqrt{\frac{2}{7}}\leq \lambda <0\lor 0<\lambda\leq 3 \sqrt{\frac{2}{7}}$\end{tabular} $\\
 \hline
\end{tabular}
    \caption{Eigenvalues and the stability conditions corresponding to each critical point for {\bf[Model-\ref{Powerlawcouplingfun}}].}
    % is used to refer to this table in the text
    \label{modelIIeigenvalues}
\end{table}
%%%%%%%

\begin{figure}[H]
    \centering
\includegraphics[width=58mm]{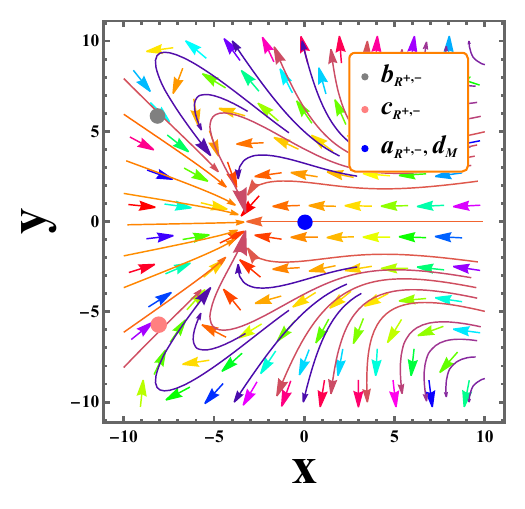}
\includegraphics[width=58mm]{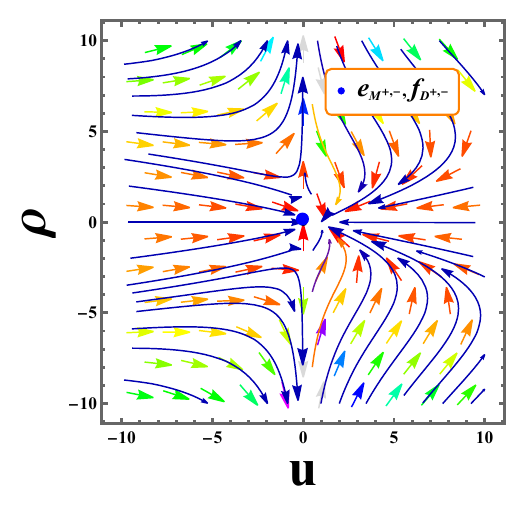}
\includegraphics[width=58mm]{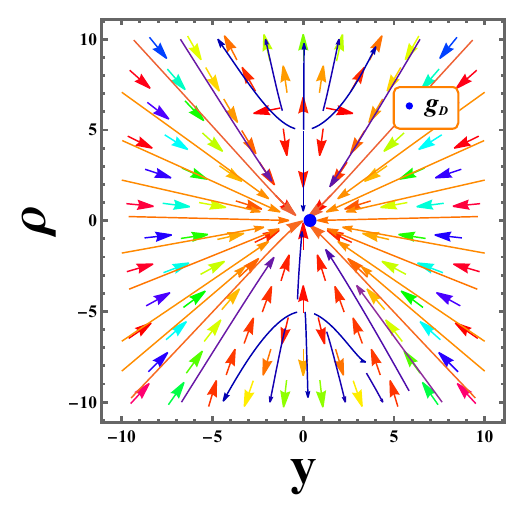}
    \caption{ 2-D Phase Portraits at $x=0,\, y=-4.9,\, u=8,\, \rho=4.5,\, \beta=-0.2, \, \beta=2$ \, (for $e_{M^{\pm}}, f_{D^{\pm}}$), \, $\beta=0$ (for $g_{D}$),$ \,   \sigma=-0.30, \, \lambda=-0.2, \, n= -1.2, \, n= 2$ (for $g_{D}$) for  [Model-\ref{Powerlawcouplingfun}].}
    \label{phasespacem1.2}
\end{figure}

\begin{figure}[H]
    \centering
\includegraphics[width=70mm]{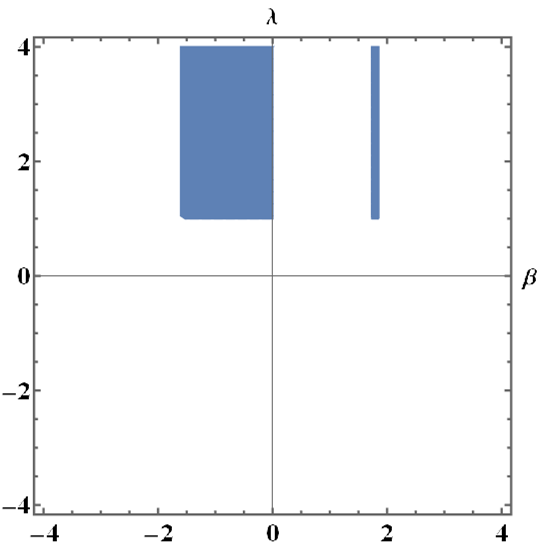}
    \caption{ 2-D Region plot for common range of stability conditions of critical points for [Model-\ref{Powerlawcouplingfun}], for$\left(-3 \sqrt{\frac{2}{7}}\leq \lambda<0\lor \sqrt{3}<\lambda\leq 2 \sqrt{\frac{6}{7}}\right) \land \beta>1$.}
    \label{2dregionplotm1.2}
\end{figure}

A detailed analysis of these critical points is presented as follows,
\begin{itemize}
\item \textbf{Radiation-Dominated Critical Points :} The critical points $a_{R^{\pm}}, b_{R^{\pm}}, c_{R^{\pm}}$ are appearing in the radiation-dominated phase of the Universe. The critical point $a_{R^{\pm}}$ is the standard radiation-dominated critical point with $\Omega_{r}=1$. This critical point has a zero, positive, and negative eigenvalue at the Jacobian matrix Ref. Table--\ref{modelIIeigenvalues} makes this critical point non-hyperbolic and is a saddle point. The phase space trajectories at this critical point are moving away; hence the saddle point behaviour can be observed from the 2-d phase portrait from Fig.--\ref{phasespacem1.2}. The other two critical points $b_{R^{\pm}}, c_{R^{\pm}}$ are the non-standard radiation-dominated critical points with $\Omega_{r}=1-\frac{4}{\lambda^2}$. These critical points are the scaling radiation-dominated solutions with $\Omega_{DE}=\frac{4}{\lambda^2}$. The eigenvalues at these critical points contain zero, positive, and negative eigenvalues; hence, these are non-hyperbolic, saddle, and unstable. The phase space trajectories are moving away from these critical points and can be observed from Fig.--\ref{phasespacem1.2}. Since the value of $\omega_{tot}=\frac{1}{3}$, hence these critical points will not describe the late-time accelerated expansion phenomena.\\
\item \textbf{Matter-Dominated Critical Points :} The critical points $d_{M}, e_{M^{\pm}}$ are the critical points that describe the matter-dominated phase of the Universe. The critical point $d_{M}$ has existence condition $2\beta n-n \ne 0$ which is same for $e_{M^{\pm}}$ but $e_{M^{\pm}}$ are described at $\beta=2$. The critical point $d_{M}$ describes a standard matter-dominated epoch with $\Omega_{m}=1$. The eigenvalues at $d_{M}$ are presented in Table-\ref{modelIIeigenvalues}, which show this critical point's non-hyperbolic saddle point behaviour. From the phase space diagram presented in Fig.--\ref{phasespacem1.2}, the trajectories at this critical point move away and show a saddle point behaviour. The critical point $e_{M^{\pm}}$ is a non-standard matter-dominated critical point with $\Omega_{m}=1-\frac{3}{\lambda^2}$. The eigenvalues at this critical point show that this is a normally hyperbolic critical point and show stability at $-2 \sqrt{\frac{6}{7}}\leq \lambda<-\sqrt{3}\lor \sqrt{3}<\lambda\leq 2 \sqrt{\frac{6}{7}}$. The phase space trajectories at this critical point show attractor behaviour can be observed from Fig.--\ref{phasespacem1.2}. Moreover $e_{M^{\pm}}$ are the scaling matter-dominated solution with $\Omega_{DE}=\frac{3}{\lambda^2}$.\\
\item{\textbf{DE-Dominated Critical Points :}} The critical points $f_{D^{\pm}}, g_{D}$ are the DE-dominated critical points. The critical points $f_{D^{\pm}}$ will exist at $2\beta n -n \ne 0, \beta=2$ and is describing standard DE-dominated critical point with $\Omega_{DE}=1$. This critical point is normally hyperbolic and shows stability at $\sqrt{3}<\lambda <0\lor 0<\lambda <\sqrt{3}$. This point can describe the late-time cosmic acceleration phenomena at $-\sqrt{2}<\lambda<\sqrt{2}$. The phase space trajectories are attracting at this critical point and can be observed from Fig.--\ref{phasespacem1.2}. The critical point $ g_{D}$ is the de-Sitter solution and can be obtained at $\beta=0, n=2$. The phase space plot is plotted for $x=0,\, y=-4.9,\, u=8,\, \rho=4.5,\, \beta=0, \,  \sigma=-0.30, \, \lambda=-0.2, n=2$ where the model parameters are constrained using the common stability range of the critical points $\left(-3 \sqrt{\frac{2}{7}}\leq \lambda<0\lor \sqrt{3}<\lambda\leq 2 \sqrt{\frac{6}{7}}\right) \land \beta>1.$ For clear visualization, we have plotted a 2-D plot for this range in Fig.--\ref{2dregionplotm1.2}. The critical point $g_{D}$ is normally hyperbolic and is showing stable behaviour within the range $-3 \sqrt{\frac{2}{7}}\leq \lambda <0\lor 0<\lambda\leq 3 \sqrt{\frac{2}{7}}$.   
\end{itemize}
%%%%%%
The common stability range for critical points $e_{M^{\pm}},f_{D^{\pm}},g_{D}$ is $\left(-3 \sqrt{\frac{2}{7}}\leq \lambda<0\lor \sqrt{3}<\lambda\leq 2 \sqrt{\frac{6}{7}}\right) \land \beta>1 $.

\begin{figure}[H]
    \centering
\includegraphics[width=70mm]{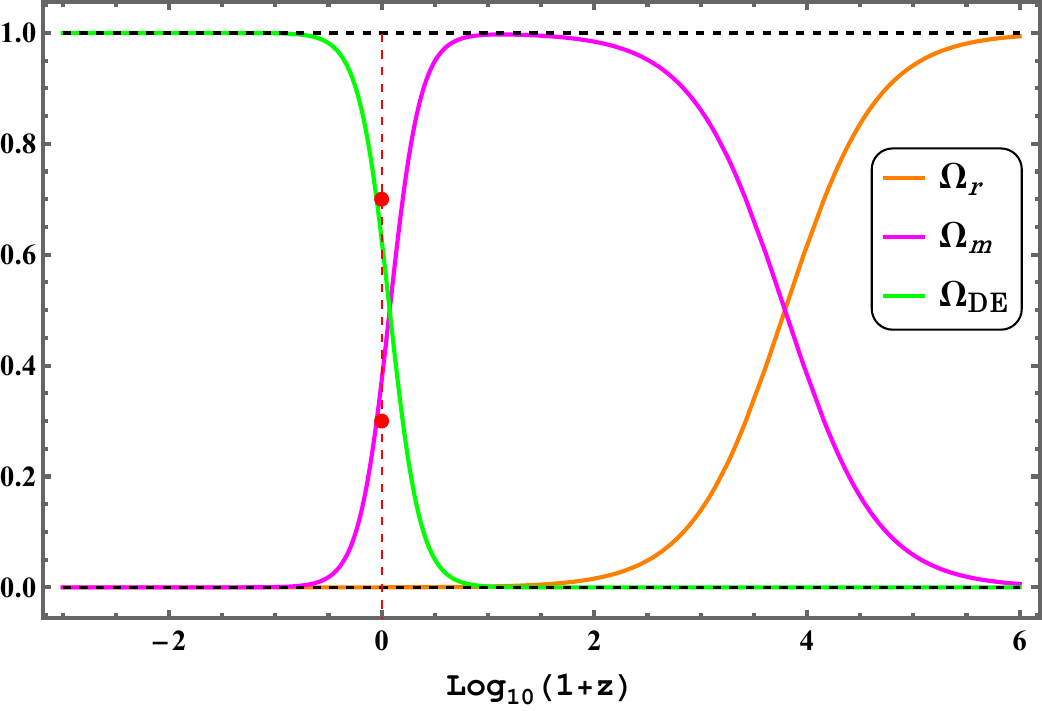}    \includegraphics[width=70mm]{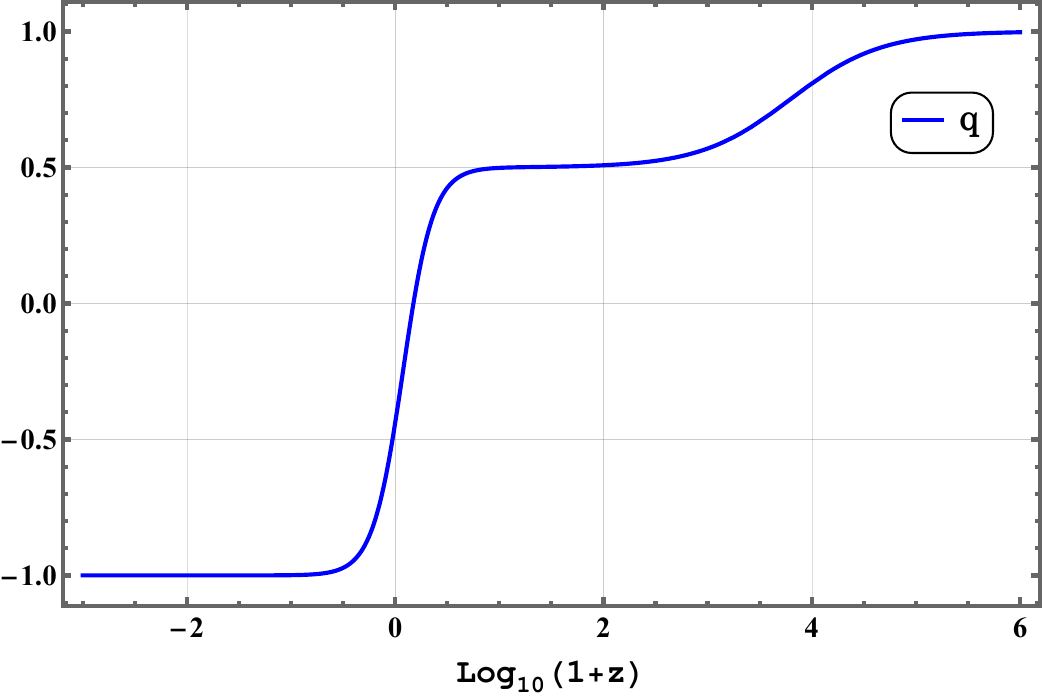}
    \caption{ {\bf Left Panel}--Density parameter for radiation $(\Omega_r)$, matter $(\Omega_m)$, DE $(\Omega_{DE})$ in redshift. {\bf Right Panel}--Deceleration parameter $(q)$ in redshift for $\beta=1.1, \,  n=0.30, \, \lambda=-0.2$ for the initial conditions $x_0=10^{-8.89}, \, y_0=10^{-2.89}, \, u_0=10^{-5.96}, \, \rho_0= 10^{-0.9}, \, \sigma_{0}=-10 \times 10^{-19}$ for {\bf[Model-\ref{Powerlawcouplingfun}}]. } \label{evolutionm1.2}
\end{figure}

\begin{figure}[H]
    \centering
\includegraphics[width=70mm]{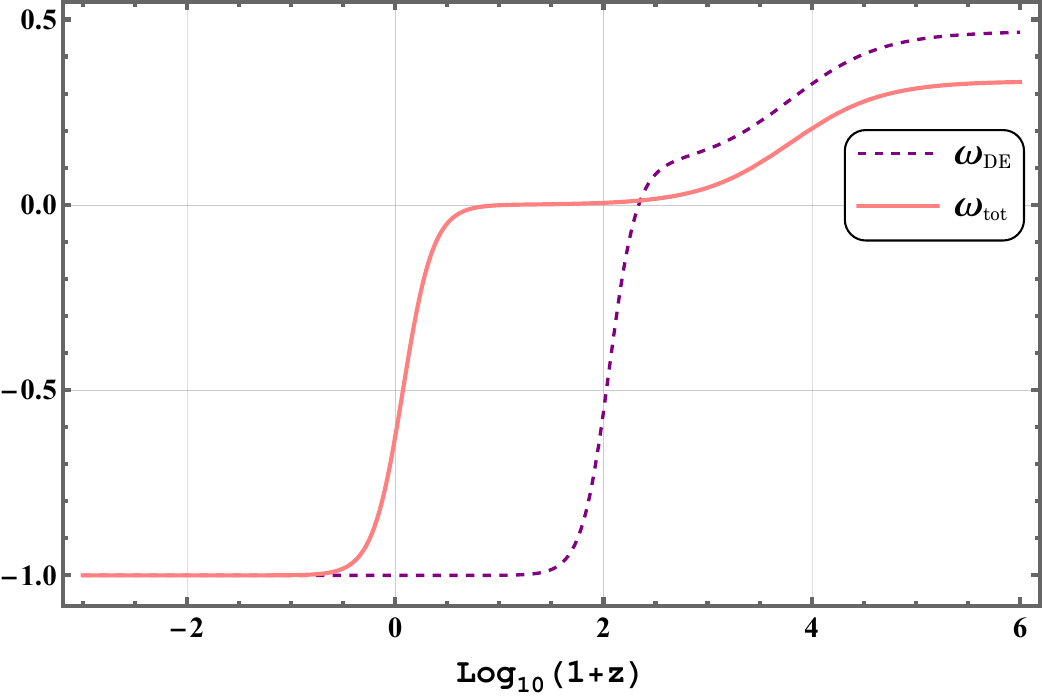}
    \caption{The DE EoS $(\omega_{DE})$parameter and total EoS $(\omega_{tot})$ parameter in redshift for $\beta=1.1, \,  n=0.30, \, \lambda=-0.2$ for the initial conditions $x_0=10^{-8.89}, \, y_0=10^{-2.89}, \, u_0=10^{-5.96}, \, \rho_0= 10^{-0.9}, \, \sigma_{0}=-10 \times 10^{-19}$ for {\bf[Model-\ref{Powerlawcouplingfun}}].} \label{Eosplotplotm1.2}
\end{figure}

%%%%%%%%%%%%%%%%%%%%%%%%%%%%%%
In Fig.--\ref{evolutionm1.2}, we have shown the evolutionary behaviour of $\Omega_{r}, \, \Omega_{m} \,, \Omega_{DE}$ from which one can conclude that at the early phase, the radiation dominated the DE and from early to late time, the DE dominates both radiation and matter. The matter-radiation equality occurred at the redshift around $z \approx 3387$ and the value of $\Omega_{DE}\approx 0.7$ \cite{Kowalski_2008} and $\Omega_{m}\approx 0.3$ \cite{Planck:2018vyg}. The transition from deceleration to acceleration occurs at the low redshift $z \approx 0.65$, which can be observed from the plot of the deceleration parameter presented in Fig.--\ref{evolutionm1.2} which is compatible with the $\Lambda$CDM model. The value of deceleration parameter at the present time is $q_{0}=-0.4944^{+0.036}_{-0.036}$ which is compatible with \cite{Capozziellomnras}. From Fig.~\ref{Eosplotplotm1.2}, the behaviour of $\omega_{DE}$ shows deceleration to acceleration from early to late time, and at present, the value of $\omega_{DE}\approx -0.993^{+0.05}_{-0.02}$ \cite{Planck:2015bue,Planck:2018vyg}.  

\section{Summary and Conclusion}\label{conclusion}
In this work, we have studied the dynamical system analysis for the recently developed scalar-torsion $f(T, \phi)$ gravity\cite{Gonzalez-Espinoza:2021mwr,Gonzalezreconstruction2021,Gonzalez-Espinoza:2020jss}  formalism. The two well-motivated non-minimally coupling functions of the scalar field $F(\phi)$, one is exponential \cite{Gonzalez-Espinoza:2020jss,roy2018dynamical} and the other one is power law form \cite{roy2018dynamical} is considered in the analysis. These forms of the coupling functions can be obtained through the reconstruction \cite{Gonzalezreconstruction2021} using the power law form of the torsion scalar $T$ in $f(T,\phi)$ gravity formalism. The autonomous dynamical system is framed to analyse critical points representing different important epochs of the Universe. The stability of these critical points is analysed using the signature of the eigenvalues of the Jacobian matrix.  \\

Model--\ref{Exponentialcouplingfun} produces seven critical points such as, $A_{R^{\pm}}, B_{R^{\pm}}, C_{R^{\pm}}$, all of them are in the radiation-dominated epoch of the Universe. The critical points show unstable behaviour, which can be seen from the $2-d$ phase space [Fig.--\ref{phasespace2dm1}]. The critical points $D_{M}, E_{M^{\pm}}$ are the matter-dominated critical points. The critical point $D_{M}$ shows saddle point behaviour and represents a standard matter-dominated era with $\Omega_{m}=1$. The critical point $E_{m^{\pm}}$ is the stable scaling matter-dominated solution and shows stability within the range $2 \sigma<\lambda\leq 2 \sqrt{\frac{6}{7}}\land \beta >\frac{\lambda-\sigma}{\lambda}$.  Moreover, the critical points $F_{D^{\pm}}, G_{D^{\pm}}$ represent the DE-dominated era of the evolutionary phase of the Universe. The DE-dominated critical points show stable behaviour and are the late-time attractors that can be visualised from the $2-d$ phase space plots presented in Fig.--\ref{phasespace2dm1}. The critical point $f_{D^{\pm}}$ describes the accelerated expansion of the Universe evolution with the range $-\sqrt{2}<\lambda<\sqrt{2}$. The exponential coupling function is capable of producing the critical points representing matter, radiation, and the DE-dominated era with the common stability range of the model parameters $-\frac{\sqrt{3}}{2}<\sigma<0\&\&\left(\beta \leq \frac{1}{2}\land -\frac{\sigma }{\beta _{10}-1}<\lambda<0\right)$, the same range can visualised from 3-d plot presented in Fig.--\ref{regionplotm1}.\\   

Model--\ref{Powerlawcouplingfun},  we have analysed the power law coupling scalar field coupling function \cite{roy2018dynamical} with the exponential scalar field potential. The additional dynamical variable $\sigma$ needs to be considered in addition to the four independent dynamical variables $x$, $y$, $u$, $\rho$. In this case, we obtain three critical points $a_{R^{\pm}}, b_{R^{\pm}}, c_{R^{\pm}}$, which represents the radiation-dominated epoch of the Universe evolution and are the unstable critical points. These critical points $d_{m}$ and $e_{M}^{\pm}$ represents the matter-dominated era with $\omega_{tot}=0$. Critical point $d_{M}$ is showing saddle point behaviour, whereas the critical point $e_{M^{\pm}}$ is a stable scaling matter-dominated solution. In this case, the critical points $f_{D^{\pm}}$ and $g_{D}$ represent the DE-dominated epoch of Universe evolution.  The critical point $f_{D^{\pm}}$ explain the accelerated expansion within the range $-\sqrt{2}<\lambda<\sqrt{2}$ and the critical point $g_{D}$ is the de-Sitter solution. We obtain the common range of the model parameters where these critical points show stability is $\left(-3 \sqrt{\frac{2}{7}}\leq \lambda<0\lor \sqrt{3}<\lambda\leq 2 \sqrt{\frac{6}{7}}\right) \land \beta>1$ and the same range can be visualised from Fig.--\ref{2dregionplotm1.2}. From this summary, we have concluded that both models are viable cosmological models and are capable of describing different important epochs of the evolution of the Universe, including the description of the late-time cosmic acceleration. This work can be extended, and these models can be validated to constrain the viable ranges of the model parameters using different recent cosmological observations.

\section*{Acknowledgement}
SAK acknowledges the financial support provided by the University Grants Commission (UGC) through the Senior Research Fellowship (UGC Ref. No.: 191620205335). BM acknowledges SERB-DST for the Mathematical Research Impact Centric Support (MATRICS)[File No : MTR/2023/000371]. SKT and BM thank IUCAA, Pune (India) for providing support for an academic visit during which a part of this work is accomplished.

\section*{References} 
\bibliographystyle{utphys}
\bibliography{references}

\end{document}